\documentclass[a4paper]{article}
\usepackage{epsfig}
\usepackage{graphicx}  
\usepackage{color}     
\usepackage{color}
\usepackage{lscape}
\usepackage{amssymb}
\date{\today}
 \normalsize

\def\be {\begin{equation}}
\def\ee {\end{equation}}
\def\bea {\begin{eqnarray}}
\def\eea {\end{eqnarray}}
\def\bc {\begin{center}}
\def\ec {\end{center}}
\def\bfg {\begin{figure}}
\def\efg {\end{figure}}
\def\bi {\begin{itemize}}
\def\ei {\end{itemize}}
\def\nn {\nonumber}

\def\no {\noindent}

\def\vs {\vspace}

\def\ie{{\it i.e.}}
\def\a  {\alpha}

\def\l  {\lambda}

\def\m  {\mu}
\def\n  {\nu}
\def\o  {\omega}

\def\r  {\rho}
\def\th {\theta}
\def\s {\sigma}

\def\beq{\begin{equation}}
\def\eeq{\end{equatfion}}
\def\br{\begin{eqnarray}}
\def\er{\end{eqnarray}}
\newcommand{\eel}[1] {\label{#1}\end{equation}}

\newcommand{\bdm}{\begin{displaymath}}
\newcommand{\edm}{\end{displaymath}}
\begin{document}
\renewcommand{\thefootnote}{\fnsymbol{footnote}}

\vspace{.3cm}

\title{\Large\bf On the correctness of cosmology from quantum potential
}

\author
{ \it \bf  E. I. Lashin$^{1,2}$\thanks{slashin@zewailcity.edu.eg},
\\
\small$^1$ Ain Shams University, Faculty of Science, Cairo 11566,
Egypt.\\
\small$^2$ Centre for Fundamental Physics, Zewail City of Science and
Technology,\\
\small Sheikh Zayed, 6 October City, 12588, Giza, Egypt.}
\maketitle

\begin{center}
\small{\bf Abstract}\\[3mm]
\end{center}
We examine in detail the cosmology based on quantal (Bohmian) trajectories as suggested in a recent study \cite{FD}.
We disagree with the conclusions regarding predicting the value of the cosmological constant $\Lambda$ and evading the big bang singularity.
Furthermore, we show that the approach of using a quantum corrected Raychaudhuri equation (QRE), as suggested in \cite{FD}, is unsatisfactory, because, essentially, it uses
the  Raychaudhuri equation, which is a kinematical equation, in order to predict dynamics. In addition, even within this inconsistent
framework, the authors have adopted unjustified assumptions and carried out incorrect steps leading to doubtful conclusions.
\\
\begin{minipage}[h]{14.0cm}
\end{minipage}
\vskip 0.3cm \hrule \vskip 0.5cm

\maketitle

It has been widely known that the cosmos, at a large scale, is to be homogenous, isotropic and spatially flat and thus could be described by Friedmann-Robertson-Walker (FRW) metric
whose line element in the comoving cartesian coordinates is given by,
\be
ds^2 =g_{\m\n}\, dx^\m\, dx^\n = -(dx^0)^2 + a(t)^2\left[(dx^1)^2 + (dx^2)^2 + (dx^3)^2\right],
\label{ds2}
\ee
where $x^\mu$ is the four dimensional coordinate, $x^\mu \equiv \left(x^0 = c\,t,\, x^1 = x,\, x^2 = y,\, x^3 = z\right)$, and $a(t)$ is the scale factor.

The metric tensor $g_{\m\n}$ can be easily read from Eq.(\ref{ds2}) to be diagonal and given by,
\be
 g_{\m\n}=\mbox{Diag}\left[-1, a(t)^2, a(t)^2, a(t)^2\right].
 \ee
The scale factor $a(t)$ can be determined by applying field equations of General Relativity (GR), namely:
\be
R_{\m\n} -{1\over 2}\, g_{\m\n} R = {8 \pi G\over c^4}\, T_{\m\n},
\label{gre}
\ee
where $R_{\m\n}$ and $R$ are the Ricci tensor and scalar respectively. As to the energy-momentum tensor $T_{\m\n}$, which in our case represents a perfect fluid having density $\rho$, pressure $p$ and velocity $U_\m$, it assumes the form
\be \label{emt}
T_{\m\n} = \left(\r + {p\over c^2}\right)U_\m U_\n + p\, g_{\m\n}.
\ee
It is worth mentioning that we are using
a metric with signature $(-, +, +, +)$, in order to be the same one used in \cite{FD}. We also do not set $c$, the speed of light, equal to unity for the sake of clarity
and to keep all fully dimensionfull factors apparent, like the Newton's gravitational constant $G$. For later manipulations, we recall that the Hubble parameter $H$ is defined to be $\dot{a}/ {a}$.

The current observations \cite{perlmutter}--\cite{bao} ranging from Type IA supernova observations,
cosmic microwave background radiation (CMBR ) data and baryon acoustic oscillations
indicate that the matter-energy content of our universe is comprised of dark energy filling $72\%$ , probably in the form of a
cosmological constant $\Lambda$ whose physical  origin is debatable,
dark matter occupying about $23\%$ whose physical non-baryonic nature is so far not determined, and the rest as an ordinary baryonic matter which could be luminous or not.

The authors in \cite{FD} used a quantum corrected Raychaudhuri equation (QRE) to derive a modified Friedmann equation which enabled them to extract
a value for the cosmological constant $\Lambda$  as well as  to evade the big-bang singularity. The QRE was obtained by replacing geodesics with quantal (Bohmian) trajectories.
In order to show why, to us, this procedure is invalid, it is appropriate to present briefly the essential ingredients for Raychaudhuri equation (RE) as can be found in many books on general relativity such as \cite{tool,nut}.

The RE is concerned as regards how a congruence of curves might evolve. The evolution can be attributed to a change of the quantity $B_{\m\n}$ which is defined as,
\be
B_{\m\n}\equiv D_\n V_\m,
\label{bdef}
\ee
where $V^\m$ is a tangent vector along the curve parameterized by affine parameter $\tau$ along the curve. For convenience, we suppress the dependence on the other parameter expressing
 how far the neighboring curves are apart from each other.  The rate of change of $B_{\m\n}$ along the curve is given as,
\be
{D B_{\m\n}\over D\tau} \equiv  V^\l D_\l B_{\m\n} = D_\n\left(V^\l D_\l V_\m\right) - \left(D_\n V^\l\right) \left(D_\l V_\m\right) -R_{\s\m\l\n} V^\l V^\s,
\label{bev}
\ee
where $D_\n$ denotes the covariant derivatives and $R_{\s\m\l\n}$ is the Riemann curvature tensor \footnote{
The convention followed for the Riemann curvature and Ricci tensors are respectively $R^\s_{\;\r\m\n}= \left(\partial_\m \Gamma^\s_{\n\r} + \Gamma^\s_{\m\a} \Gamma^\a_{\n\r}\right) -
 \left(\partial_\n \Gamma^\s_{\m\r} + \Gamma^\s_{\n\a} \Gamma^\a_{\m\r}\right)$ and $R_{\m\n} = R^\s_{\;\m\s\n}$.} Having chosen a congruence of timelike geodesics we get
 \be
{D B_{\m\n}\over D\tau} = - B^\l_{\;\;\n} B_{\m\l} -R_{\s\m\l\n} V^\l V^\s.
\label{bevs}
\ee
Defining  $\th = g^{\m\n} B_{\mu\n}$ and  plugging it into Eq.(\ref{bevs}), we get
\be
{D \th \over D\tau} = - B_{\m\n} B^{\n\m} -R_{\s\l} V^\l V^\s.
\label{tev1}
\ee
Now we decompose $B_{\m\n}$ into its irreducible parts as,
\be
B_{\m\n} = \s_{\m\n} + {1\over 3} \th P_{\m\n} + \o_{\m\n},
\label{db}
\ee
where the projector $P_{\m\n}$  projecting onto the 3-dimensional subspace orthogonal to $ V_\m$ is defined as
\be
P_{\m\n} = g_{\m\n} + {1\over \, c^2}\, V_\m V_\n.
\label{proj}
\ee
The three terms in Eq. (\ref{db}) have distinct physical interpretations in that
the trace part $\th$ describes expansion, the symmetric traceless part $\s_{\m\n} = {1\over 2} \left( B_{\m\n} + B_{\n\m}\right) -{1\over 3}\, \th\, P_{\m\n}$ describes shear and the skew symmetric part $\o_{\m\n} = {1\over 2}\, \left( B_{\m\n} - B_{\n\m}\right)$  describes rotation.

When plugging Eq.(\ref{db}) into Eq.(\ref{tev1}) we get the celebrated RE:
\be
{D \th \over D\tau} = -{1\over 3}\, \th^2  - \s_{\m\n} \s^{\m\n} +  \o_{\m\n} \o^{\m\n} -R_{\m\n} V^\m V^\n.
\label{tev2}
\ee

RE as it stands is a kinematic equation devoid of any dynamics. In order to get in touch with the physical world, the tensor $R_{\m\n}$ should be supplied
as the one that satisfies GR equations and the specified congruence must be that of timelike or null geodesics.

In a standard cosmology, we have FRW metric in the comoving frame where the cosmic fluid can be considered to be at rest $\ie$  $V^\m = (c,0,0,0)$. The quantities $\s_{\m\n}$ and $\o_{\m\n}$ are vanishing as can be easily verified, while the expansion parameter $\th$ turns out to be $3 H$. Upon using GR equations (Eq.(\ref{gre})) for $R_{00}$ we get,
 \be
 R_{00}= {4\pi G\over c^4}\left(\r c^2 + 3 p\right),
 \ee
 Plugging the result for $R_{00}$ in Eq.(\ref{tev2}) we get the `Second' Friedmann equation:
 \be
 \dot{H} = - H^2 - {4 \pi G\over 3}
\left(\r  + 3 {p \over c^2}\right).
\label{f1}
\ee
Expressing the FRW Ricci scalar as $\displaystyle R = {6\over c^2}\,\left({\ddot{a}\over a} + {{\dot{a}}^2\over a^2}\right)$ and writing $R_{00}$ from (Eqs. \ref{gre} and \ref{emt}) as,
$\displaystyle R_{00}= -{R\over 2} + {8\pi G\over c^4}\,\r c^2,$
we get the `First' Friedmann equation:
\be
H^2 = {8\pi G\over 3} \r.
\label{f2}
\ee

To summarize, we can get the two basic Friedmann equations of cosmology, Eqs.(\ref{f1},\ref{f2}), through the RE (Eq.(\ref{tev2})). However, this occurs only after imposing the GR field equations and using a congruence of timelike geodesics that coincide with the motion of the cosmic fluid. In this specific case, the expansion parameter $\th$ is obliged to be $3 H$. Thus, any obtained equation governing $H$ should be consistent with the GR field equations. Having chosen different congruences, not necessarily geodesics, then the expansion parameter $\th$ would not be $3 H$. Hence, the derived equation for $\th$ based on Eq.(\ref{bev}) should be solved to get the expression of $\th$. It is important to stress that one can not derive new dynamics from RE other than that contained in the GR field equations.

The authors of \cite{FD} have used the idea of quantal Bohmian trajectory in conjunction with RE. Although the mere idea of quantal Bohmian trajectory is a problematic one, as we shall show later, but nevertheless let us continue with our understanding of the authors' arguments. The authors started with a massive scalar field $\Phi$ coupled non-minimally to gravity. The equation of motion for $\Phi$ is taken to be in the form,
\be
\left(\Box  - {m^2 c^2\over \hbar^2} + \epsilon R \right) \Phi=0,\;\;\; \Box \equiv
g^{\m\n} D_\m D_\n ,
\label{sc1}
\ee
To get the velocity field associated with a Bohmian trajectory, one can use the polar decomposition of  $\Phi$ as,
\be
\Phi = {\cal R}\, e^{{i S\over \hbar}},
\label{pol}
\ee
then Eq.(\ref{sc1}) would imply
\bea
g^{\m\n} \left(D_\m S\right)\left(D_\n S\right)&=& -m^2 c^2 +\hbar^2 {\Box {\cal R}\over {\cal R}} +\hbar^2 \epsilon R,\nn\\
2 g^{\m\n} \left(D_\m {\cal R}\right)\left(D_\n S\right) &=& - {\cal R}\,\Box S.
\label{pole}
\eea

The velocity field $U^\m$ associated with the quantal (Bohmian) trajectory is defined to be:
\be U^\m = {D^\m S \over m} \label{Udef}\ee
and as a consequence of the first of Eqs.(\ref{pole}) we obtain,
\be
U^\m U_\m = - c^2 +{\hbar^2\over m^2} {\Box {\cal R}\over {\cal R}} +{\hbar^2\over m^2} \,\epsilon\, R.
\label{nor}
\ee
It is clear that the velocity field $U^\m$, as its norm is defined in Eq.(\ref{nor}),  is not guaranteed to be always of a timelike type. It could flip into a spacelike vector due to the presence of the terms $\displaystyle {\hbar^2\over m^2} {\Box {\cal R}\over {\cal R}}$ and $\displaystyle {\hbar^2\over m^2} \,\epsilon\, R$. This is a serious drawback that might turn the whole approach into
 being unphysical. Putting aside this difficulty, let us  continue with the usual procedure and compute the quantity $\left(D_\m U_\n\right) U^\n$ to find:
\be
\left(D_\m U_\n\right) U^\n = {\hbar^2\over 2 m^2} D_\m\!\!\left({\Box {\cal R}\over {\cal R}}\right) +{\hbar^2\over 2 m^2} \,\epsilon\, D_\m R.
\label{devg}
\ee
Using Eq.(\ref{devg}) and Eq.(\ref{bev}), one can obtain the QRE (the modified form of the RE applicable to these non-geodesic quantal trajectories) as:
\bea
{D \th \over D\tau} &=& -{1\over 3}\, \th^2  - \s_{\m\n} \s^{\m\n}  -R_{\m\n} U^\m U^\n +
{\hbar^2\over 2 m^2}\, \Box\!\!\left({\Box {\cal R}\over {\cal R}}\right) +{\hbar^2\over 2 m^2} \,\epsilon\, \Box R + A\left(\theta,\hbar,R, {\cal R}, U^\mu\right) ,\nn\\
A(\theta,\hbar,R, {\cal R}, U^\mu) &=& -\frac{2}{3} \frac{\theta}{U_\nu U^\nu} \frac{\hbar^2}{2 m^2} \left[ \epsilon U^\m D_\m R + U^\m D_\m\!\!\left({\Box {\cal R}\over {\cal R}}\right) \right],
\label{modR1}
\eea
where $\tau$ is the parameter defined by $\displaystyle U^\m = \frac{dx^\m}{d\tau}$.
Let us compare the QRE of (Eq. \ref{modR1}) with the corresponding QRE in \cite{FD}:
\bea
\label{QRE1}
{D \th \over D\tau} &=& -{1\over 3}\, \th^2  -R_{\m\n} U^\m U^\n +
{\hbar^2\over  m^2}\, P^{\m\n}D_\m D_\n \left({\Box {\cal R}\over {\cal R}}\right) +{\hbar^2\over  m^2} \,\epsilon\, P^{\m\n}D_\m D_\n R.
\eea

We find that the two QRE equations do not coincide in that the $\sigma^2$ and the $A$ terms in (Eq. \ref{modR1}) are absent in (Eq. \ref{QRE1}), and the numerical factors
appearing in front of $\hbar^2$ are also different even though  the precise value
of these factors are not relevant for our intended considerations. We trace the absence of the $A-$term in (Eq. \ref{QRE1}) to the `improper' definition in \cite{FD} of
the trace part
\be \tilde{\theta} = P^{\m\n} D_\n U_\m \ee
in that it uses the projector $P^{\m\n}$ as defined in (eq. \ref{proj}) instead of using the FRW- metric $g^{\m\n}$
for the correct definition of the trace part: \be \theta = g^{\m\n} D_\n U_\m .\ee This explains also the form of the two $\hbar^2$-terms in (Eq. \ref{QRE1}) and why they differ from
the corresponding terms in (Eq. \ref{modR1}). We note here, considering the new velocity squared-norm (Eq. \ref{nor}), that the definition of
the projector should be amended to be:
\be
P_{\m\n} = g_{\m\n} - {1\over \, U^\a U_\a}\, U_\m U_\n.
\label{projnew}
\ee

 We note however that using either $P^{\m\n}$ or $g^{\m\n}$  would amount in \cite{FD} to the same result for the $\hbar^2$-terms in Eqs (\ref{modR1}, \ref{QRE1}) upon neglecting higher orders of $\hbar$. This
comes because the velocity in \cite{FD} was defined as $\displaystyle U^\m = \hbar \frac{D^\m(\frac{S}{\hbar})}{m} \equiv \hbar \frac{D^\m\tilde{S}}{m}$ and so
the $U^\m U^\n$-term
in the `imprecise' definition of the projector (Eq. \ref{proj}), in \cite{FD}, was considered as a quantum correction producing a higher order term in $\hbar$ (this argument would not apply for the $A$-term since the $P^{\m\n}$-
contribution vanishes and one needs to compute explicitly the $U^\m U^\n$-contribution in order to evaluate the full $g^{\m\n}$-contribution).

As to the term containing $\s_{\m\n}$, describing shear, one should not discard it for the specific choice of the velocity field $U^\m$ with norm given by Eq.(\ref{nor}). Actually the non-zero components for $\s_{\m\n}$ turn out to be dependent on both $H$ and $S$. When assuming $S$ depending only on $t$ in order to be consistent with homogeneity and isotropy, then
the non-zero components for $\s_{\m\n}$ can be easily computed and we get:
\be
\s_{00}  =  {\ddot{S}\over 3 m c^2}\,\left(2 + {\dot{S}^2\over m^2 c^4}\right) + H\,\left( {\dot{S}^3 \over m^3 c^6} - {\dot{S}\over m c^2}\right),\;\;
\s_{11} = \s_{22} = \s_{33} = {\ddot{S} a^2 \over 3 m c^2}.
\ee

Again, putting aside the discrepancies in the two QREs, the authors of \cite{FD} have later manipulated their QRE (Eq. \ref{QRE1}) confusingly, to us, several times:
 \begin{description}
\item{First:}
They have identified $\th$ with $3\,H$ which is inexact, in our opinion, because the particle motion described by the velocity field $U^\m$, as given in Eq.(\ref{nor}),  is not a geodesic motion. Assuming $S$ to be $t$-dependent in accordance with homogeneity and isotropy, then $\th$ is given by $\displaystyle \th = -{3\over m c^2}\, H\, \dot{S} -{\ddot{S}\over m c^2}$. Thus, the identification of $\th$ with $3 H$ and interpreting  the resulting equation for $H$ as a new Friedmann equation is not accurate.
\item{Second:} They have replaced $R_{\s\l} U^\l U^\s$ by $\displaystyle \frac{4 \pi G}{c^2} (\rho c^2 + 3 p)$ which is not completely true as it should be replaced by $\displaystyle \frac{4 \pi G}{c^4} (\rho c^2 + 3 p) \frac{\dot{S}^2}{m^2}$ for a $t$-dependent $S$.
\item{Third:}
They have assumed a value for ${\cal R}$ equal to $\exp{\left(-r^2/L_0^2\right)}$ where $r=\sqrt{x^2 + y^2 + z^2}$, and they considered $L_0$ as the present size of the universe. The position dependence of ${\cal R}$  is not in accordance with homogeneity and isotropy of the FRW-metric, as the latter two features admit only time dependence for any scalar function.
\item{Fourth:}
They have identified the contribution $\displaystyle {\hbar^2\over 2 m^2}\,\Box\! \left({\Box {\cal R}\over {\cal R}}\right)$ as a cosmological constant whose value would be $\displaystyle \Lambda_Q = {\hbar^2\over 2 m^2 c^2}\,\Box\!\left({\Box {\cal R}\over {\cal R}}\right)$ which is not a constant as stated in \cite{FD}. The value of $\Lambda_Q$ turns out to be,
\be
\Lambda_Q = {\hbar^2\over 2 m^2 a^4 L_0^4 c^2}\left[ 6 + {a \ddot{a}\over c^2}\left(2 r^2 - 3 L_0^2\right)\right].
\ee
\item{Fifth:}
They have carried out the calculations with ordinary derivatives instead of covariant derivatives, especially in computing $\Lambda_Q$ which explains how they got a diffrent result ($\frac{1}{L_0^2}$). Apparently, the authors in \cite{FD,sd} considered using the spacetime metric as a fixed background which would allow them to use ordinary derivatives instead of covariant ones. We think this is not correct because one will lose the general covariance.
\end{description}

Leaving aside all the above mentioned debatable points, which are enough to put into doubt the work carried out in \cite{FD}, let us, for the sake of complete judgement, continue the discussion
of the QRE obtained in \cite{FD}. After having identified
$\th$ with $3 H$,  $R_{\s\l} U^\l U^\s$ with $\frac{4 \pi G}{c^2} (\rho c^2 + 3 p)$, and after having dropped the $\s^2$ and $A$-terms in (Eq. \ref{modR1}), and after having discarded
 the term which was identified as a cosmological constant (-fortunately not relevant for the foregoing discussion-) we get, assuming the validity of the GR field equations and using
 the equation of state $p = \omega\, \r\, c^2$, the following equation:
\be
\dot{H} = -{3\over 2} \left(1+ \omega\right) H^2 + {\hbar^2\over 6 m^2} \,\epsilon\, \Box R.
\label{modR2}
\ee
Apart from an irrelevant, for our discussion, numerical factor in front of $\hbar^2$,  Eq.(\ref{modR2}) coincides with the corresponding equation in \cite{FD}.
Now, regardless of any quantum theory of gravitation having GR as a zeroth-order approximation, one can substitute, up to a multiplicative numerical factor, $T$ -the trace of the energy-momentum tensor- for $R$ in Eq. (\ref{modR2}) as the difference would constitute a higher order in $\hbar$. Therefore, one can guess certain basic features that should be present in any contribution  proportional to $\Box R$, namely:
  \begin{description}
\item{First:}
 It should vanish for the case of radiation where $(\omega = {1\over 3})$ since
 $T$  is zero for radiation.
\item{Second:}
It should vanish for the case of dark energy (cosmological constant where $\omega = -1$) since $T$ is constant in this case.
\item{Third:}
It should vanish for the case of $\omega = -{1\over 3}$ that corresponds to the case of Milne universe which is equivalent to an empty universe dominated by negative curvature term.
\end{description}

However, assuming the GR field equations (which lead to $\displaystyle R=-{8 \pi G\over c^4} T$ where $ T = g^{\m\n} T_{\m\n} =-\r c^2 \left(1-3 \omega\right)$), and the
continuity equation ($\dot{\r} = - 3 H \left(1+\omega\right) \r$) and
the Friedmann equation ($\dot{H}=-{3\over 2} \left(1+ \omega\right) H^2$), one can evaluate the contribution proportional to $\Box R$ in Eq. (\ref{modR2}) to get:
\be
\dot{H} = -{3\over 2} \left(1+ \omega\right) H^2 -{9\hbar^2\over 4 m^2 c^4} \,\epsilon\, \left(1 - 9\, \omega^2\right) \left(1 + \omega\right) H^4.
\label{modR3}
\ee
which disagrees with what was obtained in \cite{FD}:
\be
\dot{H} = -{3\over 2} \left(1+ \omega\right) H^2 -{6\hbar^2\over m^2 c^4} \,\epsilon\,\left(1 + \omega\right) \left[6\left(1 + \omega\right)^2 -{81\over 2} \left(1 + \omega\right) + 18\right] H^4.
\label{modR4}
\ee

The result in Eq.(\ref{modR3}) possesses all the required features while that of Eq.(\ref{modR4}) does only posses the second feature. The correct equation (Eq.\ref{modR3}) reduces exactly to
the GR equation in case of radiation, which is the most relevant case for the early universe. Having no deviation from GR in case of radiation-dominated universe
 implies that if we trace the cosmic evolution backward in time we would hit the big-bang singularity in a finite time. This finding does not agree with the conclusion derived in \cite{FD} based on the equation (\ref{modR4}) where the big-bang singularity is completely evaded.

Avoiding the big-bang singularity can be simply explained through interpreting the equations governing $H$, like Eq.(\ref{modR3}) or Eq.(\ref{modR4}), as a dynamical system. The fixed points associated with the dynamical system, where $\dot{H}=0$, can determine qualitatively the dynamical behaviour of $H$. For more elaborations on the notion of dynamical system and fixed points in a one dimensional flow, one can consult \cite{strog}.

Creating attractive fixed points in the past at a finite value of $H$ prevents $H$ from running away to infinity in a finite time (a situation characterizing a big bang singularity). The system, instead, will be directed towards the fixed point in an infinite time lapse. The equation (\ref{modR3}) ceases to create these attractive fixed point in the past for the case of radiation dominated universe ($\omega = {1\over 3}$). The sign of the contribution proportional to $H^4$ in Eq.(\ref{modR3}), responsible for creating non-trivial fixed points,  is controlled by the factor $\epsilon\, \left(1 - 9\, \omega^2\right) \left(1 + \omega\right)$. This factor is zero for $\omega = {1\over 3}\; \mbox{or}\; -1$, which are the most relevant cases for the early universe.

In more details, one can examine, for positive epsilon, the relative sign of both contributions that are proportional to $H^2$ and $H^4$.  The interesting values  for $\omega$ should lie in the interval  $[-1, {1\over 3}]$ which contain the mostly motivated physical cases. When $\omega \in ]-{1\over 3}, {1\over 3}[$, the terms proportional to $H^2$ and $H^4$ have the same sign and thus no possible fixed points other than $H=0$ can be found.
On the other hand, the opposite is true for $\omega \in ]-1, -{1\over 3}[$ where the two contributions are against each other and a fixed point other than $H=0$ can be generated.
Regarding the case of negative $\epsilon$, the consequences concerning the two intervals  $\omega \in ]-{1\over 3}, {1\over 3}[$  and   $\omega \in ]-1, -{1\over 3}[$ are swapped with each other. To sum it up,
we did not find a deviation from GR when $\omega = {1\over 3}\; \mbox{or}\; -1$, and these two cases are very relevant for the early universe. In less relevant cases for the early universe, where $\omega \neq {1\over 3}\; \mbox{or}\; -1$, one can have a deviation from GR which is not so significant.

A serious persistent drawback for all values of $\omega$, linked to the velocity field $U^\m$ with norm defined in Eq.(\ref{nor}), is that it can not be guaranteed to be always timelike. The value of the squared-norm of $U^\m$ in the case ${\cal R} = \exp{\left(-r^2/L_0^2\right)}$ is given by
\be
U^\m U_\m = - c^2 +\frac{\hbar^2}{m^2}\left({ 4 r^2 - 6 L_0^2 \over a^2 L_0^4}\right) +{3\hbar^2\over m^2 c^2} \,\epsilon\,\left(1-3 \omega\right) H^2,
\label{nor2}
\ee
It is clear, now, that it is difficult to tune the vector type to be timelike through the whole  history and in different patches of the universe. It suffices
to take a large value for $r$ in order to enforce $U^\m$ to be spacelike.

We conclude that the work carried out in \cite{FD} is questionable, particularly over the following two points:
\begin{itemize}
\item
 Extracting dynamics from a kinematical relation like RE in conjunction with an unlikely identification of the parameter $\th$ and suspect calculations leading to problematic
 conclusions.
\item
The velocity field $U^\m$ associated with the quantal (Bohmian) trajectory can not be guaranteed to be always timelike, and this difficulty alone is sufficient to make the whole approach unphysical.
\end{itemize}
Hence, in our opinion, the conclusions presented in \cite{FD}, concerning evading the big-bang singularity and predicting a value for the cosmological constant $\Lambda$, are disputable.

\vs{.2cm}
\no {\bf Acknowledgment} We thank  N. Chamoun for many useful discussions.

\no

%



\end{document}